\documentclass[sigconf]{acmart} 
\usepackage[utf8]{inputenc}
\usepackage{xcolor}
\usepackage{verbatim}
\usepackage{cleveref}
\usepackage{xspace}
\usepackage{graphicx}
\usepackage{caption}
\usepackage{subcaption}
\usepackage{censor}
\usepackage{listings}
\usepackage{lipsum}
\usepackage{xcolor}
\usepackage{soul}
\sethlcolor{black}
\makeatletter
\newif\if@blind
\@blindtrue 
\if@blind \sethlcolor{black}\else
   
\fi

\acmConference[PREPRINT]{}{}{}

\newcommand{\code}[1]{\texttt{#1}}
\newcommand{\earlybird}{early-bird\xspace}
\newcommand{\Earlybird}{Early-bird\xspace}
\newcommand{\minimd}{MiniMD\xspace}
\newcommand{\minife}{MiniFE\xspace}
\newcommand{\miniqmc}{MiniQMC\xspace}

\title{Measuring Thread Timing to Assess the Feasibility of \Earlybird Message Delivery}

\author{W. Pepper Marts}
\email{wmarts@sandia.gov}
\affiliation{
  \institution{Sandia National Laboratories}
  \city{Albuquerque}
  \state{New Mexico}
  \country{USA}
}

\author{Matthew G. F. Dosanjh}
\email{mdosanj@sandia.gov}
\affiliation{%
  \institution{Sandia National Laboratories}
  \city{Albuquerque}
  \state{New Mexico}
  \country{USA}
}

\author{Whit Schonbein}
\email{wwschon@sandia.gov}
\affiliation{%
  \institution{Sandia National Laboratories}
  \city{Albuquerque}
  \state{New Mexico}
  \country{USA}
}

\author{Scott Levy}
\email{sllevy@sandia.gov}
\affiliation{%
  \institution{Sandia National Laboratories}
  \city{Albuquerque}
  \state{New Mexico}
  \country{USA}
}

\author{Patrick G. Bridges}
\email{bridges@unm.edu}
\affiliation{%
  \institution{University of New Mexico}
  \city{Albuquerque}
  \state{New Mexico}
  \country{USA}
}

\date{July 2022}

\usepackage{natbib}
\usepackage{graphicx}

\begin{document}

\begin{abstract}

\Earlybird communication is a communication/computation overlap technique that combines fine-grained 
communication with partitioned communication to improve application run-time. Communication is divided
among the compute threads such that each individual thread can initiate transmission of its portion of the data as soon as it is complete rather than waiting for all of the threads.  
However, the benefit of \earlybird communication depends on the completion timing of the individual threads.  

In this paper, we measure and evaluate the potential overlap, the idle time each thread experiences
between finishing their computation and the final thread finishing. These measurements help us understand whether a given 
application could benefit from \earlybird communication. We present our technique for gathering this data and 
evaluate data collected from three proxy applications: \minife, \minimd, and \miniqmc. To characterize the behavior of these workloads,
we study the thread timings at both a macro level, i.e., across all threads across all runs of an application, 
and a micro level, i.e., within a single process of a single run. We observe that these applications exhibit significantly 
different behavior. While \minife and \miniqmc appear to be well-suited for \earlybird communication because of their wider thread distribution and more frequent laggard threads, the behavior of \minimd may limit its ability to leverage \earlybird communication.

\end{abstract}

\maketitle

\section{Introduction}
\label{sec:int}

To deal with the increased network demands of exascale supercomputers, many approaches have been proposed to support fine-grained communication in applications~\cite{dinan2013enabling,grant2019finepoints}.
Recent work explores the performance effects of utilizing different multithreaded communication models~\cite{marts2021minimod} as well as the  distribution of times when threads complete their work~\cite{yiltan2022}. 
Existing work has assumed that thread arrival times (the relative time at which threads rejoin at the end of a parallel compute section) follow a normal distribution, and that there regularly laggard threads (thread arrival times that are significantly later than the mean arrival time). However, previous work has not used empirical data to characterizing thread arrival distributions.

In this paper, we instrument and profile three proxy applications to capture the distribution of their thread arrival 
times.  A statistical analysis of the resulting data creates a real world characterization of threaded HPC communication, and advances our understanding of how thread arrival times may change over the course of an application run. 
By providing a better understanding of these factors, the results reported in this paper provide an important empirical basis for the evaluation of different communication interfaces and multi-threaded communication models, such as partitioned communication.

The contributions of this paper are:
\begin{itemize}
    \item A methodology for evaluating application thread behavior for multithreaded communication models;
    \item A study of three proxy applications to identify thread arrival distributions; and 
    \item An analysis of the applicability of \earlybird communication given the arrival distributions in three important HPC proxy applications.
\end{itemize}

The rest of this paper is structured as follows. 
Section~\ref{sec:bac} explains the background and the problem the paper addresses.
Section~\ref{sec:ins} discusses the instrumentation and experimental set-up for this paper.
Section~\ref{sec:res} presents the results of our experiments. 
Section~\ref{sec:dis} discusses the implications of this work and presents our plan for future extensions. 
Section~\ref{sec:rel} contextualizes our work in the body of related work.
Section~\ref{sec:con} concludes our paper.

\section{Background}
\label{sec:bac}

HPC applications traditionally adopt a bulk synchronous processing (BSP) model, where computation phases alternate with communication phases. In the BSP model, 
multithreading is often limited to the computation phase because communication between threads in different processes can incur significant overheads~\cite{thakur2009test}. Regardless, application developers continue to express a desire to use multithreaded communication~\cite{bernholdt2020survey}, and a variety of solutions to the problem have been explored, including optimizing message matching~\cite{flajslik2016mitigating}, employing software offloading~\cite{vaidyanathan2015improving}, and using partitioned communication~\cite{grant2019finepoints}. 

Partitioned communication is a strategy for dividing a communication buffer into smaller pieces such that communication can be independently initiated from different execution contexts (e.g., threads).  Determining how to divide communication buffers and when to initiate communication is a complex, application-dependent question.  For the purposes of the discussion in this paper, we use a simple model whereby each thread is assigned an equal, contiguous portion of the communication buffer and is responsible for initiating transmission of its portion of the data.   

Currently, the most prominent approach for partitioned communication is described in the MPI 4.0 standard.  However, the concept of partitioned communication is not limited to MPI.  Therefore, the data and discussion in this paper are intended to apply to partitioned communication broadly rather than to MPI's definition of partitioned communication specifically.

\begin{figure}[h!]
\centering
\includegraphics[width=\columnwidth]{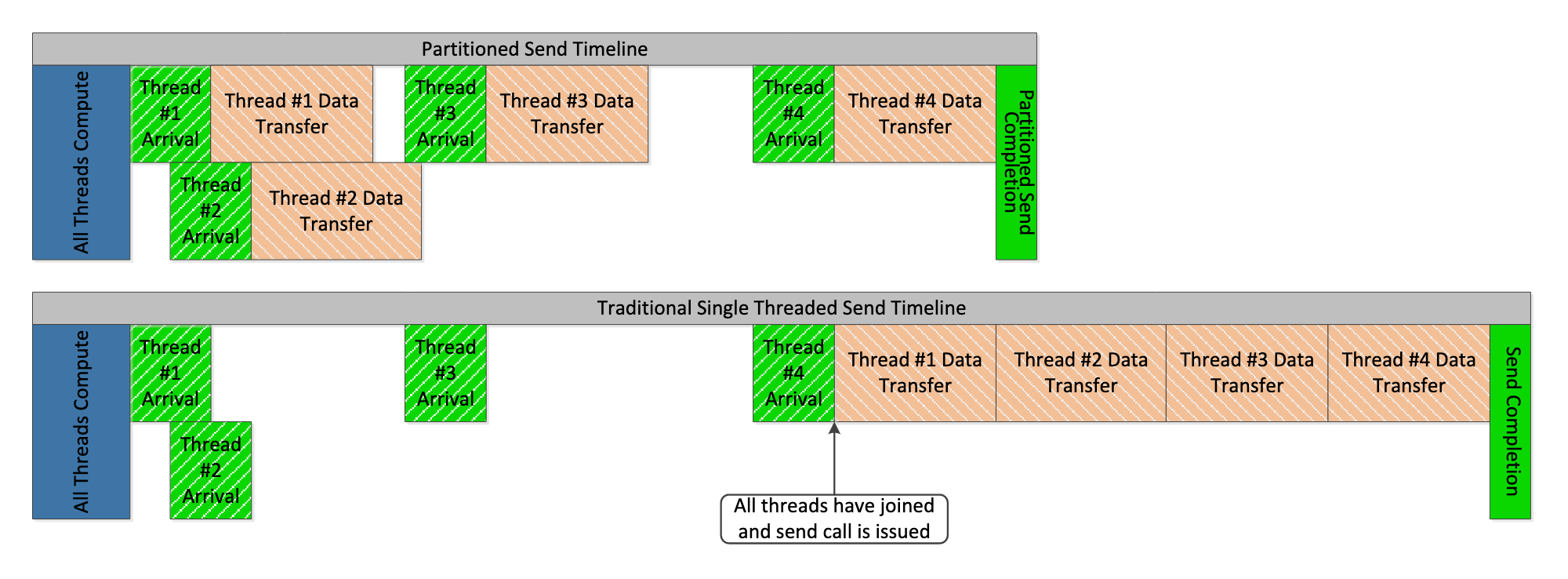}
\caption{Early-bird model of communication.}
\label{fig:earlybird}
\end{figure}

One benefit of efficient multithreaded communication is that programmers can move communication calls to be near compute, increasing network utilization. This creates a form of communication/computation overlap called ``\earlybird communication''. \Cref{fig:earlybird} provides an illustration of how \earlybird communication for partitioned communication in our model might work. 

\begin{figure}[h!]
\centering
\includegraphics[width=\columnwidth]{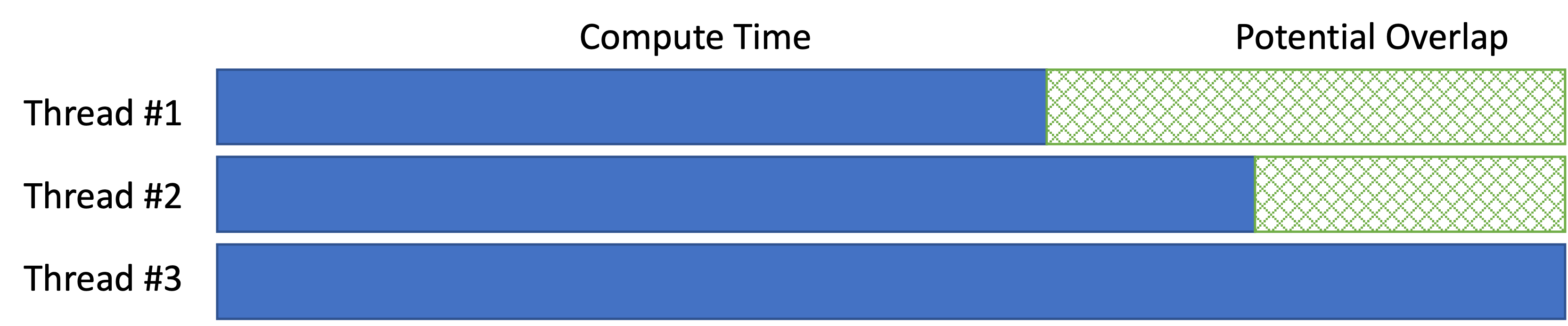}
\caption{Potential for computation-communication overlap.}
\label{fig:overlap}
\end{figure}

This leaves the question: how much do thread arrival times vary? If the thread arrival times are too similar, we expect applications to see a negative performance impact from moving to partitioned communication. 
However, as thread arrival times increase in variability the time for \earlybird communication increases. Distribution likely has an impact as well; if an application is waiting on a single laggard thread (such as in OS noise~\cite{morari2011quantitative}) it may be able to complete the transmission of all but the data produced by the laggard thread before the laggard thread finishes its computation. \Cref{fig:overlap} illustrates this with green boxes representing the time available for potential overlap. 

\section{Instrumentation and Experimental Design}
\label{sec:ins}
\subsection{Proxy Applications}
\label{sec:ins:proxy}

We instrumented three proxy applications, \minimd, \minife, and \miniqmc~\cite{heroux2011mantevo} in order to characterize the distribution of thread arrival times in a threaded compute section. 
In particular, we measure the times at which each thread enters and exits various OpenMP parallel-for regions for each iteration on each process. Timing data was collected using \code{clock\_gettime()} with \code{CLOCK\_MONOTONIC}, as defined in the IEEE POSIX.1-2017 standard \cite{posix}. The standard guarantees that the returned value is the time in nanoseconds since an undefined event in the past and that on a given core the time returned by \code{clock\_gettime()} is never earlier than a previous call to the same function. 

The POSIX standard does not provide this ordering guarantee across an entire multithreaded process spread across multiple cores and sockets. The necessary synchronization is not standard, and its existence is indicated by the \code{tsc\_reliable} CPU flag. This feature is not available on many systems, including our test platform. In order to convert the returned value into a form that is comparable outside of a given compute core, we instead calculate a derived data point, \emph{compute time}, as shown in \Cref{fig:overlap}. Compute time for each thread is elapsed time in nanoseconds, and is calculated as the difference between the time the thread exits the parallel region and the time it enters. This subtraction cancels out any divergence in the results and allows for the comparison of thread timings across cores, sockets, and nodes. 

\begin{lstlisting}[ caption={Example of code instrumentation for data collection. i is iteration.}, label=code:instrument, captionpos=b]
#pragma omp parallel
{
  int t = omp_get_thread_num();
  #pragma omp barrier
  clock_gettime(CLOCK_MONOTONIC, &t_start[i][t]);
  #pragma omp for nowait
  for(int n = 0; n < nsteps; i++) {
    // do work
  }
  clock_gettime(CLOCK_MONOTONIC, &t_end[i][t]);
  #pragma omp barrier
}
\end{lstlisting}

As illustrated in Listing~\ref{code:instrument}, in each measured compute region, we nested the main \code{\#pragma omp for} loop (adding the nowait flag) of each compute section inside of a \code{\#pragma omp parallel} region. 
This allowed us to efficiently collect start and stop times in a threaded context without consideration of internal loop indexing. As we are using elapsed time as an estimate of thread arrival time, we add a \code{\#pragma omp barrier} before the collection of thread start times to synchronize the threads.

\subsection{Experimental System and Application Configuration}

Data was collected on the Manzano cluster. 
Each node has two 24-core Intel Cascade Lake CPUs running at 2.90 GHz and 192 GB of RAM. The machine uses the RHEL7 operating system and runs on an Intel Omni-Path network. Data collected on this system used OpenMPI 4.1.1 and all executables were compiled with GCC version 10.2.1. Each application was run for ten trials with eight processes per job. Each job used all 48 available hardware thread contexts, and was configured to run for two hundred iterations. For each process and each iteration, the application gathers timing data for each of its $48$ threads.

For \minimd~\cite{heroux2011mantevo}, a parallel molecular dynamics proxy application based on LAM\-MPS~\cite{thompson_lammps_2022}, we timed all threaded compute sections in the application. Data shown is from the Lennard-Jones forcing function, the most computationally intensive section of the application. Data was collected with a compute volume of $ 128 ^ 3 $. For \minife, an unstructured mesh finite element solver, 
we timed the matrix vector product: the linear algebra function of highest order. Data was collected with a compute volume of $ 200 ^ 3 $ matrix elements per process.
For \miniqmc, a quantum Monte Carlo proxy application based on QMCPACK \cite{kim2018}, we timed the entirety of the computation for the individual threaded "movers". Although \miniqmc does not do meaningful inter-process communication, the class of applications it represents often perform considerable inter-node communication and \miniqmc serves as proxy the threading behavior we measure in this paper.

\section{Results}
\label{sec:res}
This section presents analysis of thread arrival times for each application. In order to evaluate the the sources of potential communication computation overlap put forward in \Cref{sec:int}, we break our analysis into two sections. First, an analysis of the potential normality of thread arival times as aggregated at three scales, and second, an analysis of laggard thread arrivals and characterization of classes of thread arrival distribution.

\subsection{Evaluation of Thread Arrival Times for Normality}
\label{sec:normality}

This section explores three logical groupings for understanding thread arrival time distributions: (1) At the level of an entire application across all trials and processes; (2) thread timings aggregated at the level of application iteration (the iteration count used by an application for a time step across all processes); and (3) aggregating at the level of individual processes on a single iteration (one processes' thread pool for a parallel compute region). We will refer to these as \emph{application level aggregation}, \emph{application iteration level aggregation}, and \emph{process iteration level aggregation}, respectively. Existing work in the study of early-bird communication often assumes that thread arrival can be modeled based on a single distribution. We test this assertion for our three selected groupings. 

Application level aggregation is explored for the possibility that thread arrival times can be described by a single normal distribution. 
In order to determine the normality 
of our thread arrival time distributions, we performed three tests for each application: D’Agostino~\cite{dagostino1971}, Shapiro-Wilk~\cite{shapiro1965}, and Anderson-Darling~\cite{stephens1974}. Data presented for Anderson-Darling is for a significance level of 5\%. Each test assumes a null hypothesis that the data is normally distributed. 
Each test was run on the complete data set of application level thread arrival times for a total of $768000$ samples per application. Results for all three of our applications led to rejecting the null hypothesis 
that the data is from a normal distribution. This strongly suggests that using a single, normal distribution of thread arrival times for every rank, trial, and iteration in each application is not valid model. 

Thread timing data allows determining how application behavior can vary over the course of program execution. The thread arrival times for individual application iterations can be tested to see if they can be described by a normal distribution on an iteration by iteration basis. Running the same tests for each of the $200$ application iterations that contain $3840$ samples from each application resulted in identical results for \minife and \minimd: thread arrival times for individual application iterations are not normally distributed. For \miniqmc however, there were eight application iterations for which D'Agostino's test failed to reject the null hypothesis. These same eight iterations did reject the null hypothesis for both Shapiro-Wilk or Anderson-Darling. It does not appear valid to assume that thread arrival times for an individual application iteration are normally distributed.

\begin{table}[htbp]
\centering
\begin{tabular}{|c|c|c|c|}
\hline
Test & \minife & \minimd & \miniqmc \\
\hline
D’Agostino & $3\%$ & $77\%$ & $95\%$\\
Shapiro-Wilk & $<1\%$ & $74\%$ & $96\%$\\
Anderson-Darling &  $<1\%$ & $76\%$ & $96\%$\\
\hline
\end{tabular}
\caption{Process iteration normality test results. Each cell is the percentage of aggregated process iterations that passed the normality test (i.e. failed to reject the null hypothesis). }
\label{tab:proc-iter}
\end{table}

Finally, testing at the finest explored level of aggregation, the process iteration, shows how threads join in an individual process' parallel compute context. Tests are run for each application on each for the $16000$ process iteration level sets that contain $48$ thread arrival samples. \Cref{tab:proc-iter} presents the results of these tests. For the majority of \minimd and \miniqmc process iterations, arrival times were normally distributed. For \minife less than $3\%$ of process iterations were normally distributed. We see that the normality of process iteration arrival times varies depending on the application, with our three applications demonstrating the classes of nearly completely normal, nearly completely non-normal, and a mix of normal and non-normal. Process iteration arrival times for some applications can be modeled with a normal distribution, but it is not a constant.

\begin{figure}
\centering
\begin{subfigure}{\columnwidth}
  \centering
  \includegraphics[width=\columnwidth]{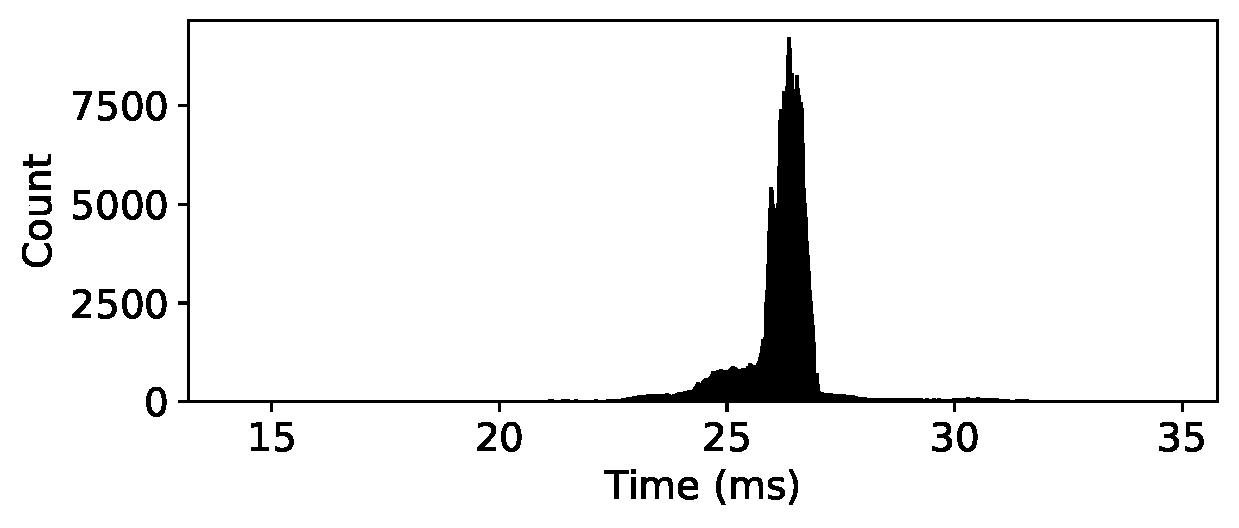}
  \caption{\minife}
  \label{fig:app_hist_sub1}
\end{subfigure}\\
\begin{subfigure}{\columnwidth}
  \centering
  \includegraphics[width=\columnwidth]{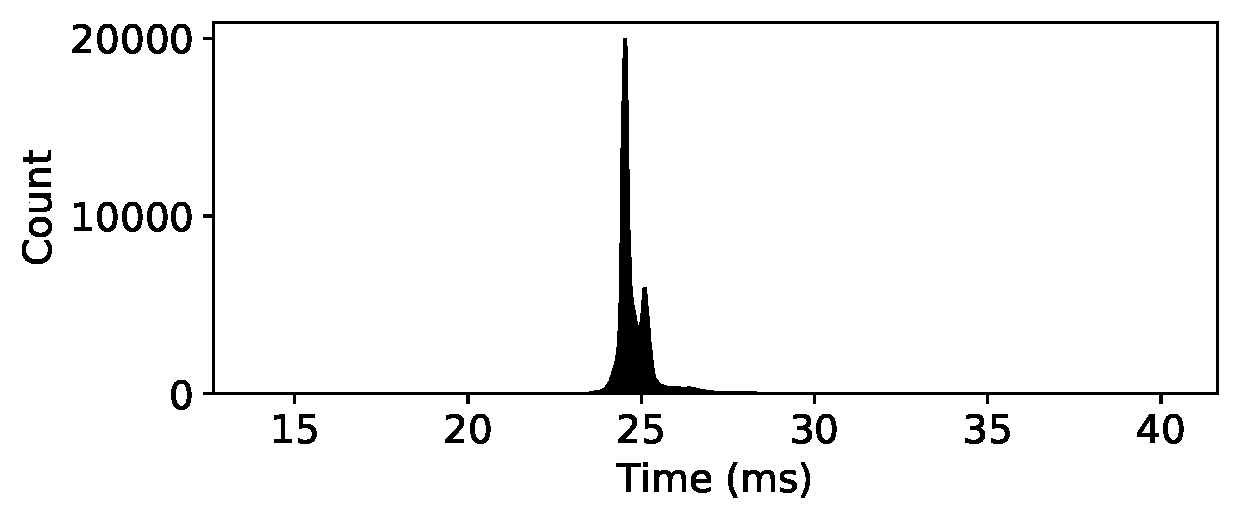}
  \caption{\minimd}
  \label{fig:app_hist_sub2}
\end{subfigure}\\
\begin{subfigure}{\columnwidth}
  \centering
  \includegraphics[width=\columnwidth]{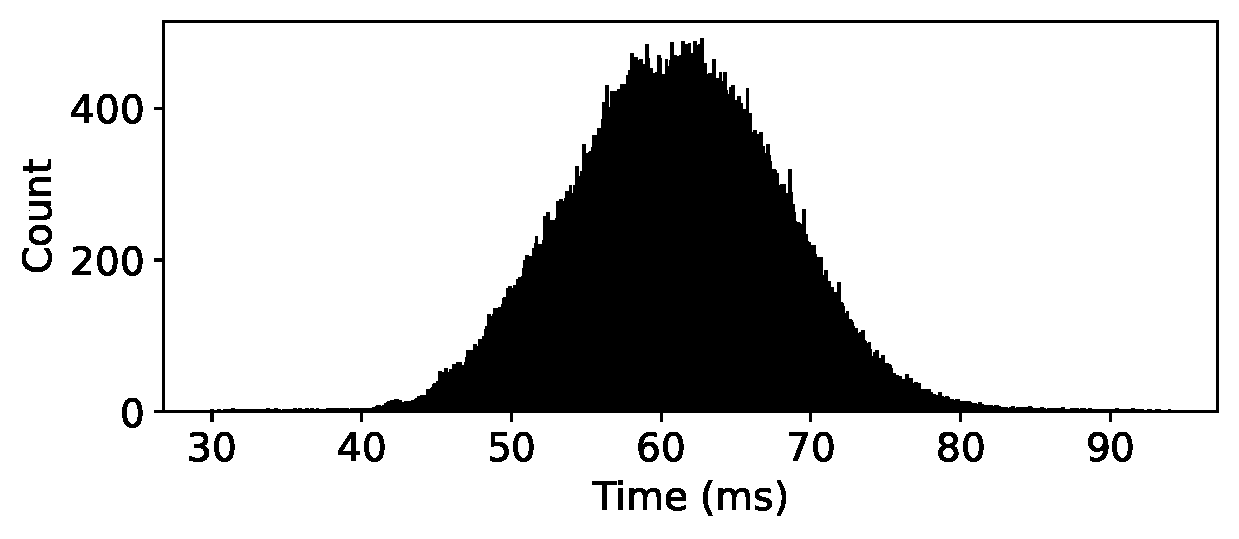}
  \caption{\miniqmc}
  \label{fig:app_hist_sub3}
\end{subfigure}
\caption{Application thread arrival time histograms for each of our three applications. Each has a bin width of 10 microseconds.
}
\label{fig:app_hist}
\end{figure}

Even for applications where individual process iterations can be modeled with normal distributions, their aggregation in application level thread arrival time behavior remains complex. We have observed large variations both within and between applications. This indicates that no single distribution is representative for all applications at any our examined aggregation levels. \Cref{fig:app_hist} presents histograms of cumulative application level thread arrival times for each of our three applications with a bin width of 10 microseconds. In the \Cref{sec:laggard} we characterize the particulars of each these application in greater detail.

\subsection{Analysis of Laggard Thread Arrivals and Reclaimable Time}
\label{sec:laggard}

In this section we present the thread arrival distributions of the three applications, with a focus laggard thread arrivals and high-level trends across application iteration. Percentile plots display all thread arrival timings across each process and trial for a total of $3840$ samples per iteration. Histograms in this section provide examples of arrival patterns observed within a process. The data presented in each histogram is only from a single iteration, process, and trial and is used to describe the time available for early-bird communication. The histograms are used to typify patterns seen in multiple iterations. Times shown are in milliseconds and represent time spent in the measured parallel compute region not the overall time spent in that iteration. Reclaimable time was determined by the summing the difference between the latest thread in that process iteration and each preceding thread. This paper presents two metrics. The first is the average amount of reclaimable time per iteration as averaged over the entire data set. The second is the average proportion of the time spent idle that iteration which is computed as the ratio between the cumulative time spent idle by all threads that iteration and the latest arrival time that iteration multiplied by number of threads. We will refer to these as \emph{average reclaimable time} and the \emph{ratio of time spent idle}, respectively.

\subsubsection{\minife}

\begin{figure}[h!]
\centering
\includegraphics[width=\columnwidth]{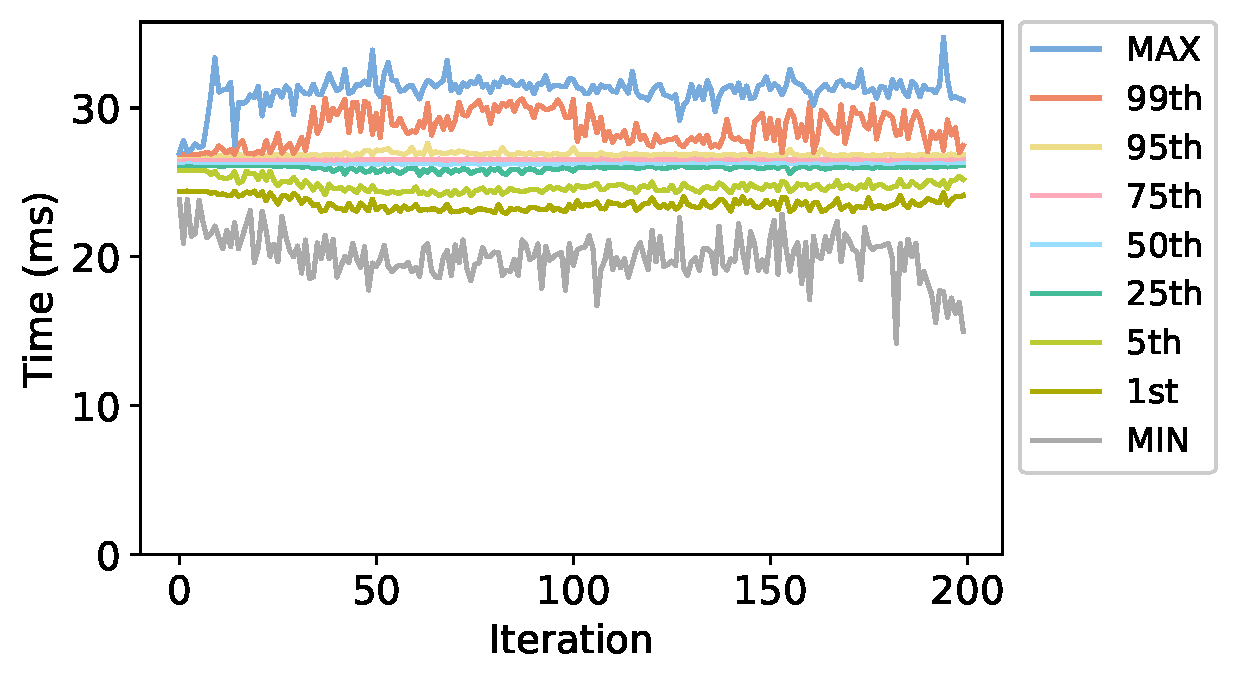}
\caption{\minife: Time spent in computing matrix-vector product.  Legend values correspond to percentiles of the collected thread execution times.}
\label{fig:minife_run_iter}
\end{figure}

\Cref{fig:minife_run_iter} presents the thread arrival times for \minife as a percentile plot. The inter-quartile range has an average value of $0.18ms$ and a maximum value of $4.24ms$. Based on \cref{fig:app_hist_sub1} and  that the 5th and 25th percentiles are generally further from the median than the 95th and 75th percentiles, we can see that early arrival is significantly more common than late arrival for this application. 
 The early threads are potentially due to work distribution imbalance; an outer loop iterates over 200 planes of the problem space and are distributed to 48 threads. 

\begin{figure}
\centering
\begin{subfigure}{\columnwidth}
  \centering
  \includegraphics[width=\columnwidth]{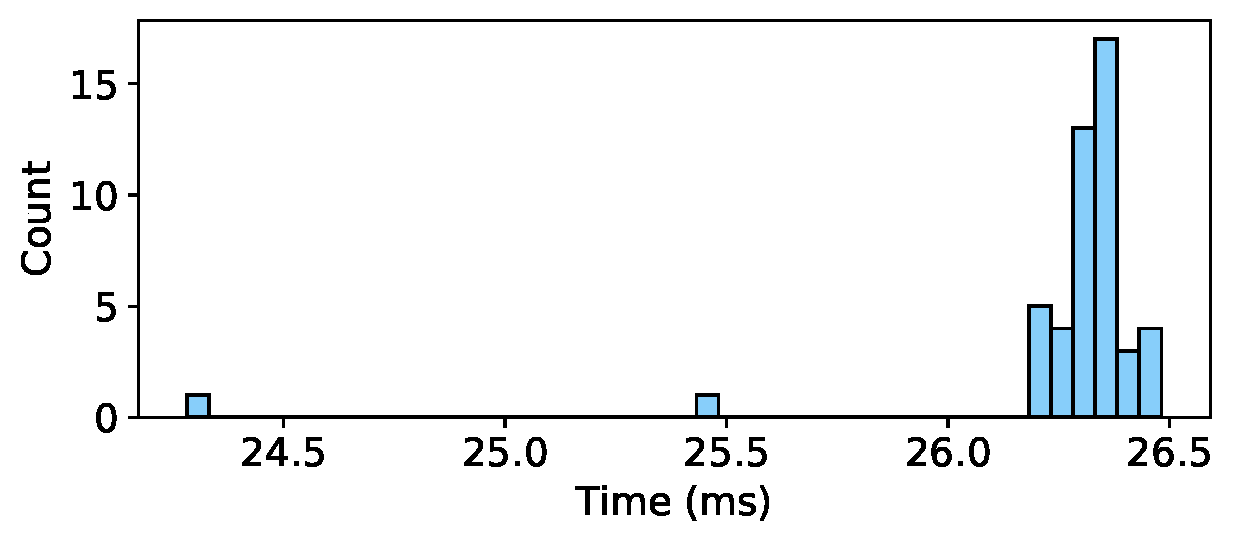}
  \caption{$77.6\%$ of recorded iterations contain no laggard thread}
  \label{fig:minife_hist_sub1}
\end{subfigure}\\
\begin{subfigure}{\columnwidth}
  \centering
  \includegraphics[width=\columnwidth]{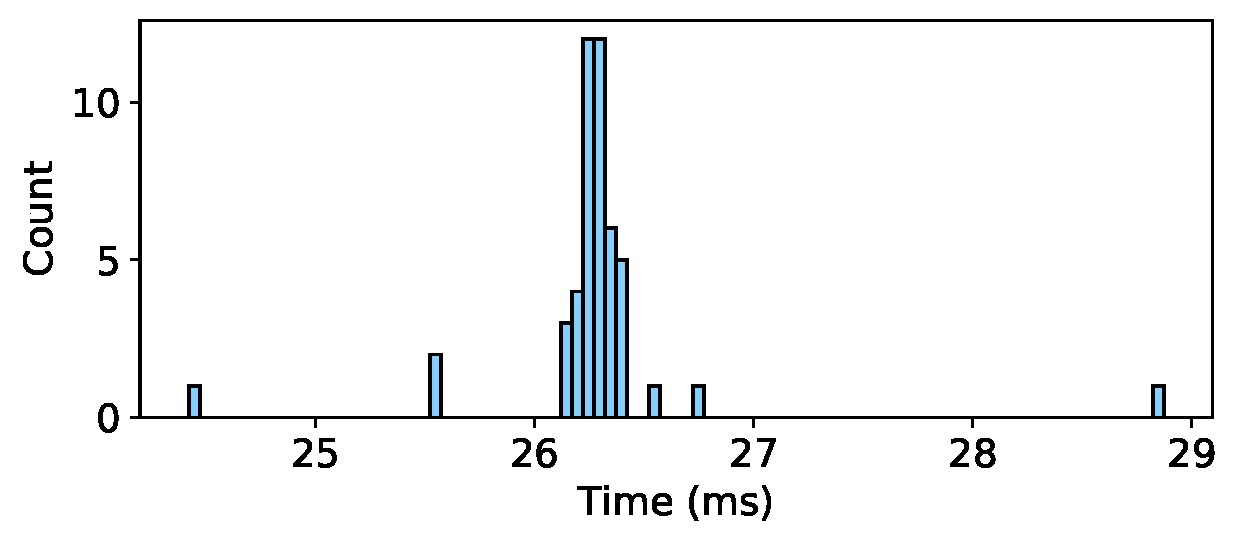}
  \caption{$22.4\%$ of recorded iterations contain a laggard thread.}
  \label{fig:minife_hist_sub2}
\end{subfigure}
\caption{Example histograms of \minife thread arrival distribution classes.}
\label{fig:minife_hist}
\end{figure}

\Cref{fig:minife_hist} shows a pair of histograms presenting thread completion times taken from our \minife data. 
Each bin has a width of $50 \mu s$. In order to determine the reclaimable time that could potentially be used for early-bird communication, we examined the difference between the arrival of each thread and the last thread to arrive during that iteration on that process. We observed two patterns, \Cref{fig:minife_hist_sub1} shows a pattern without a laggard, and \Cref{fig:minife_hist_sub2} has a clear laggard thread. To identify how many of the observed iterations had a laggard, we found the difference between the median and maximum thread time and compared that to a threshold of $1 ms$. 
This value was chosen in order to determine if a thread arrival was approximately $5\%$ slower than the mean median thread.
We determined that only in $22.4\%$ of iterations was the latest thread to arrive more that $1 ms$ slower than the median thread. Regardless of whether a laggard was present, we can see a very tight distribution of thread arrivals. The mean median thread arrival time is $26.30ms$. This corresponds to the peak in \Cref{fig:app_hist_sub1}. 
Previous experiments reveal that the distributions in \Cref{fig:minife_hist} are not normally distributed, but barring outliers they are symmetric and unimodal. The average reclaimable time was $42.82 ms$ with a $0.1928$ ratio of time spent idle.

\subsubsection{\minimd}

\begin{figure}[h!]
\centering
\includegraphics[width=\columnwidth]{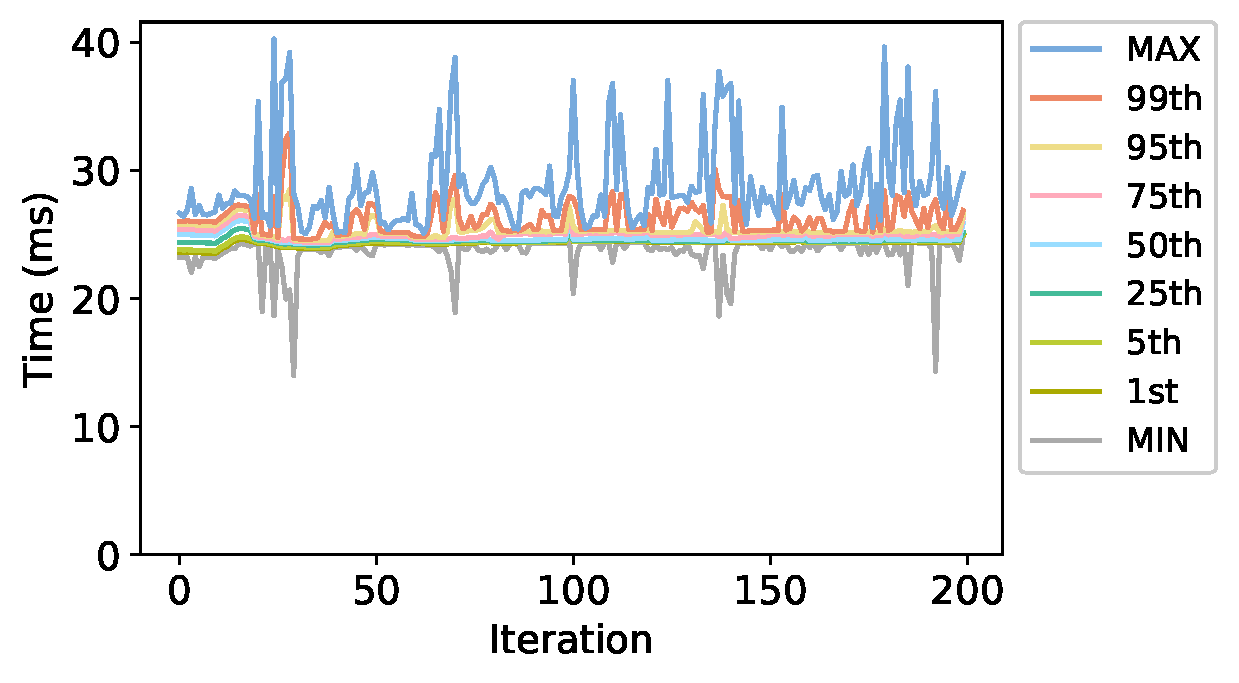}
\caption{\minimd: Time spent in Lennard-Jones forcing function.  Legend values correspond to percentiles of the collected thread execution times.}
\label{fig:minimd_run_iter}
\end{figure}

\Cref{fig:minimd_run_iter} presents the thread arrival times for \minimd as a percentile plot. These results show that two very different thread arrival distribution behaviors occur across application iteration. For the first nineteen iterations there is a significantly wider distribution of thread arrivals, which differs from the remainder of the application. This initial section appears to have consistent distribution of arrival times and few outliers of significant magnitude. This is followed by a section with sporadic laggard threads and extremely few early arrivals. The inter-quartile range for the first section has an average value of $0.93ms$ and a maximum value of $1.45ms$ while the inter-quartile range for the second section has a much lower average value of $0.15ms$ and a much higher maximum value of $7.43ms$.

\begin{figure}
\centering
\begin{subfigure}{\columnwidth}
  \centering
  \includegraphics[width=\columnwidth]{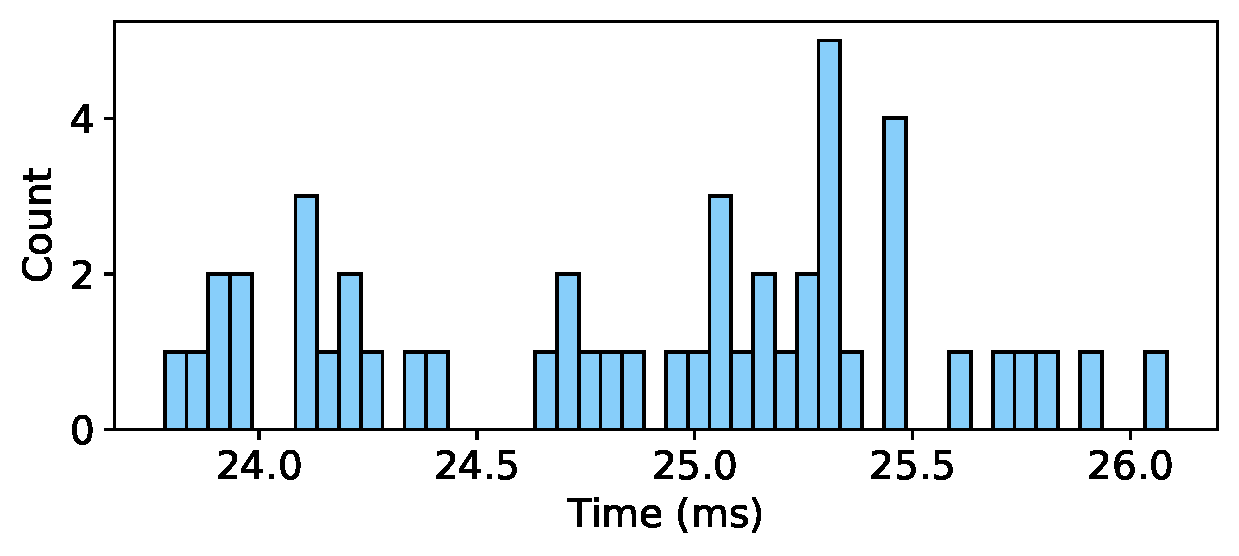}
  \caption{Initial behavior (iterations one through nineteen).}
  \label{fig:minimd_hist_sub1}
\end{subfigure}\\
\begin{subfigure}{\columnwidth}
  \centering
  \includegraphics[width=\columnwidth]{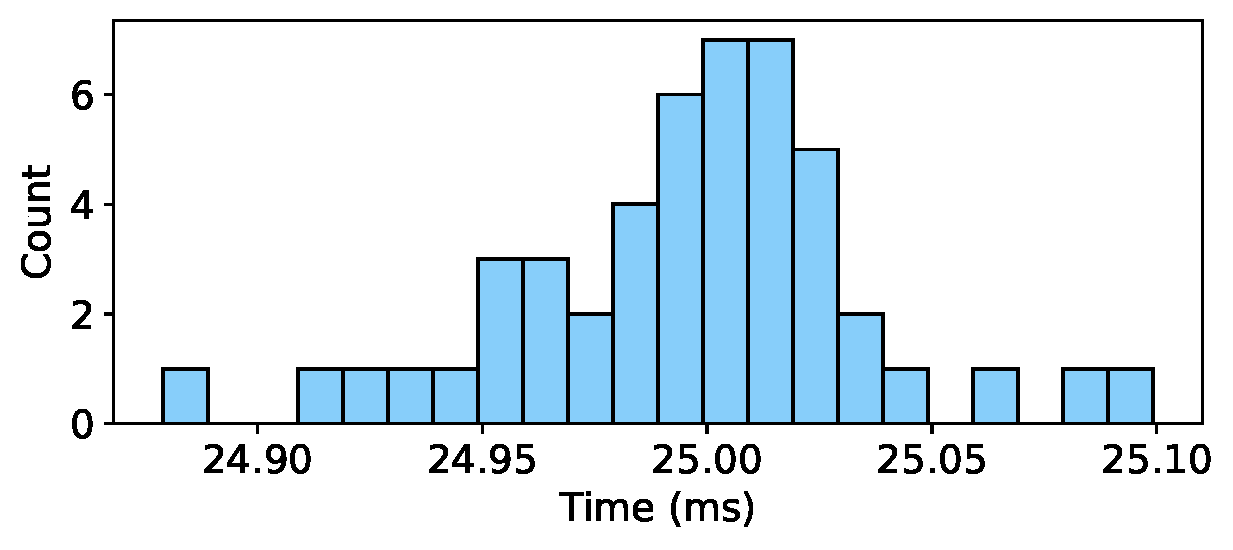}
  \caption{$95.2\%$ of recorded iterations contain no laggard thread.}
  \label{fig:minimd_hist_sub2}
\end{subfigure}\\
\begin{subfigure}{\columnwidth}
  \centering
  \includegraphics[width=\columnwidth]{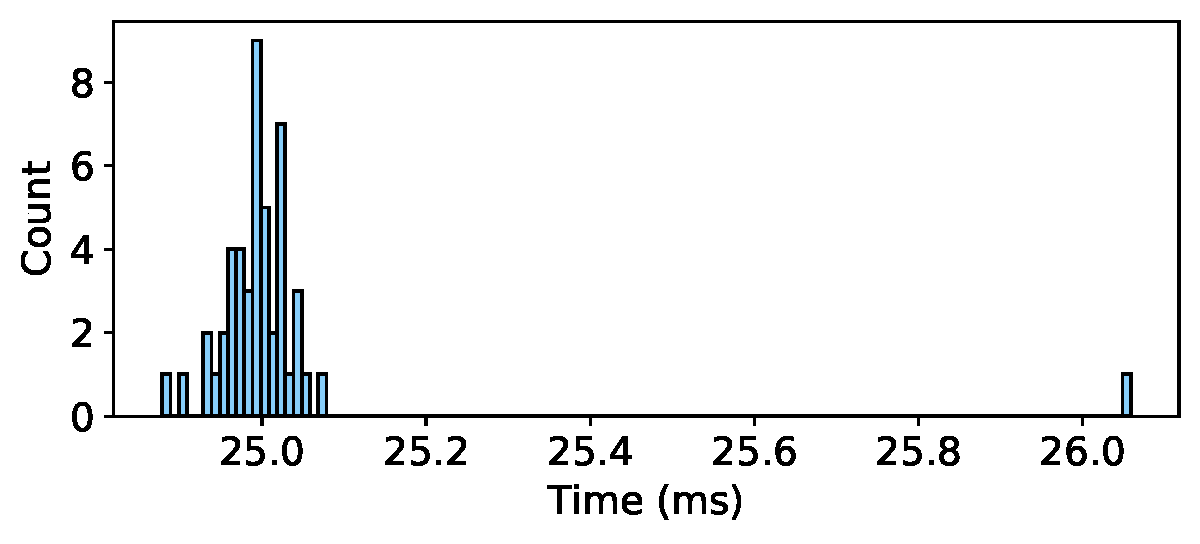}
  \caption{$4.8\%$ of recorded iterations contain a laggard thread.}
  \label{fig:minimd_hist_sub3}
\end{subfigure}
\caption{Example histograms of \minimd thread arrival distribution classes.}
\label{fig:minimd_hist}
\end{figure}

\Cref{fig:minimd_hist} shows three histograms, each representative of a subset of the observed distributions. \Cref{fig:minimd_hist_sub1} gives an example of a distribution from the first nineteen iterations. Each bin has a width of $50 \mu s$. We found that the spread of times seen in the percentile plots is not a result of variation in process or trial but is instead present in individual iterations. The observed distributions were highly consistent, with a range of just over $2ms$ a median of between $25ms$ and $26ms$. \Cref{fig:minimd_hist_sub2,fig:minimd_hist_sub3} provide examples of the remainder of computation. Each bin has a width of $10 \mu s$. We determined that only in $4.8\%$ of iterations was the latest tread to arrive more that $1 ms$ slower than the median thread. Regardless of whether a laggard was present, we can see a very tight, normal distribution of thread arrivals. The mean median thread arrival time is $24.74ms$. This corresponds to the peak in \Cref{fig:app_hist_sub2}. The average reclaimable time was $17.61 ms$ with a $0.5012$ ratio of time spent idle.

\subsubsection{\miniqmc}

\begin{figure}[h!]
\centering
\includegraphics[width=\columnwidth]{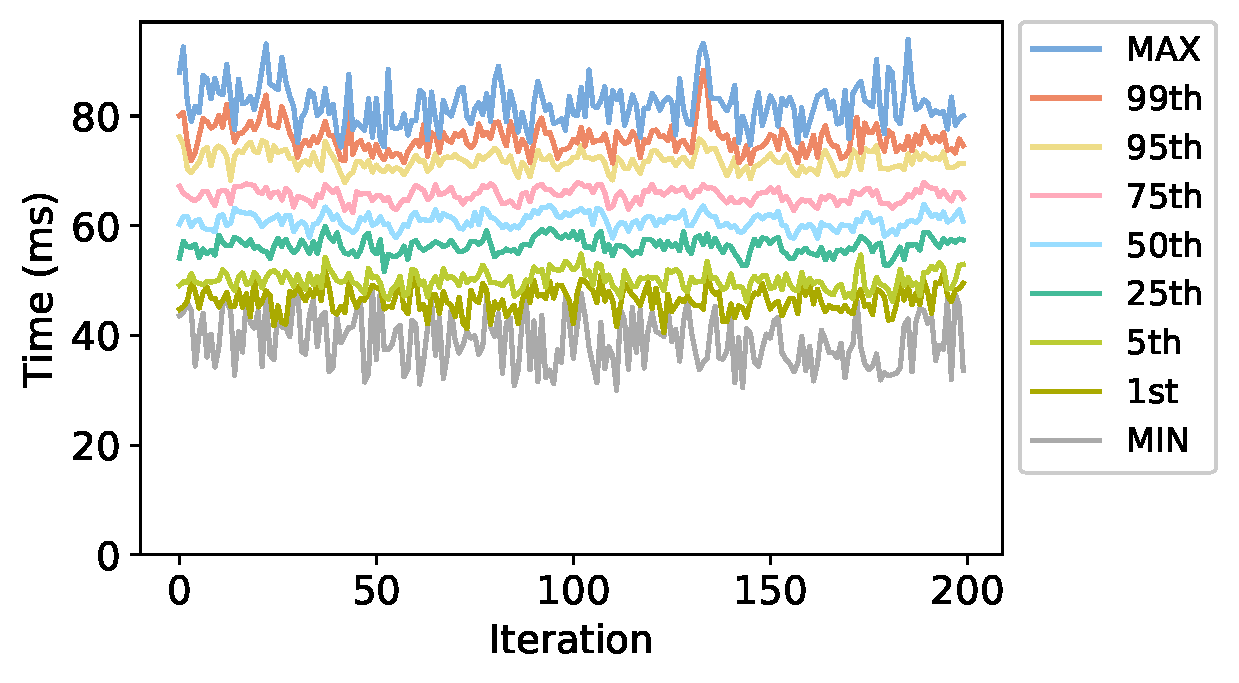}
\vspace{-15pt}
\caption{\miniqmc: Time spent computing movers.  Legend values correspond to percentiles of the collected thread execution times.}
\label{fig:miniqmc_run_iter}
\end{figure}

\Cref{fig:miniqmc_run_iter} presents the thread arrival times for \miniqmc as a percentile plot. These results show that \miniqmc has the most uniform thread arrival distribution of the applications tested. Previous experiments reveal that these distributions are normally distributed. There is little variation across iterations. It is also notable for having the highest magnitude of variation among thread arrival times during each iteration with the inter-quartile range having a maximum value of $15.61ms$ and a mean value of $9.05ms$.

\begin{figure}[h!]
\centering
\includegraphics[width=\columnwidth]{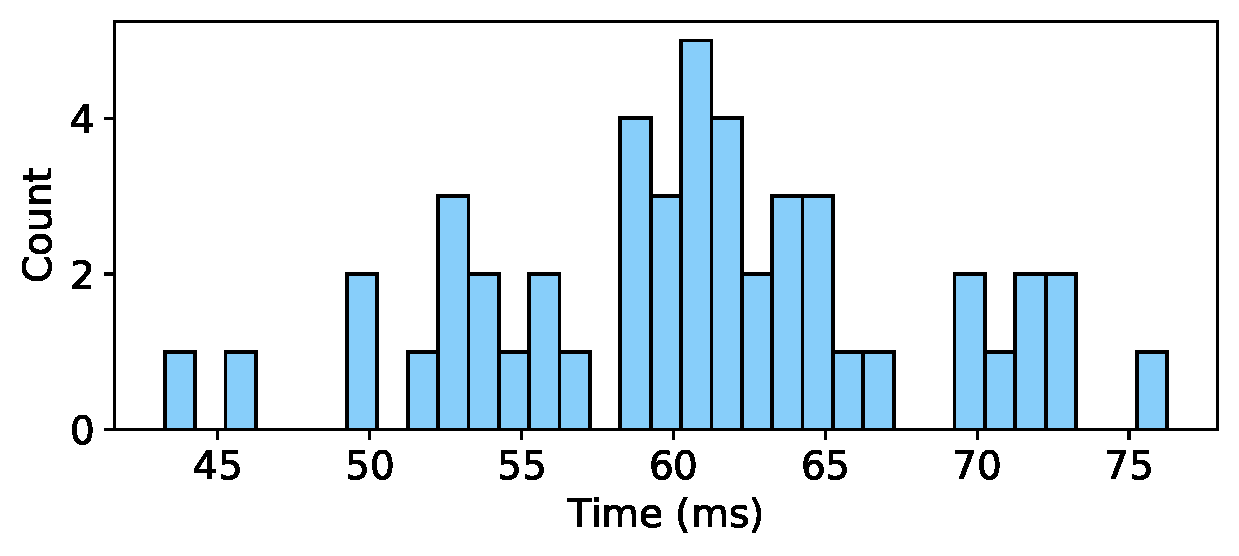}
\caption{Example histogram of \miniqmc thread arrival distribution.}
\label{fig:miniqmc_hist}
\end{figure}

In order to determine whether the individual iterations have such a wide range of run-times or if the broad distributions are from the aggregation of the 80 process trial pairs, \Cref{fig:miniqmc_hist} presents an example distribution from our \miniqmc data. Each bin has a width of $1ms$. Our data showed that the breadth of over $40ms$ in the observed arrival times present in \Cref{fig:miniqmc_run_iter} are the result of variation of thread arrivals in each iteration. The mean median thread arrival time is $60.91ms$. This corresponds to the peak in \Cref{fig:app_hist_sub3}. The average reclaimable time was $708.03 ms$ with a $0.5033$ ratio of time spent idle.

\section{Discussion and Future Work}
\label{sec:dis}
The reclaimable time results in the previous section demonstrate significant fork/join idle times that could be leveraged by \earlybird communication in a restructured application to improve application performance.  First, our analysis shows distributions of thread arrival times that have a large variance and 
significant numbers of laggard threads that suggest that \earlybird transmission may improve communication performance.  While these features are not present in every application or with perfect consistency, each application evaluated exhibited at least one of these two features frequently. Specifically, \minimd and \miniqmc show significant thread idle times (50.12\% and 50.33\%, respectively) due to laggard threads (\minimd), and the large variance of thread arrival times (\miniqmc). While \minife exhibited lower idle times (19.2791\%), 22\% of \minife iterations included laggard threads that could be potentially exploited by \earlybird communication. 

This highlights the opportunity to significantly improve application performance by taking advantage of this idle time for \earlybird communication. 
However, successfully doing so would likely require significant changes to the applications, for example fusing multiple existing fork/join loops that precede communications or changing the overall communication plans of the applications. Current applications are not structured to do so, which is why our approach has focused on measuring extant fork/join idle times in the applications to understand the scale of this opportunity. Given the sample of proxy applications in this paper we see abundant opportunity for \earlybird communication and reaffirm the assumptions of the existing literature.

This data also provides insight into potential directions for best implementing \earlybird communication in applications, an active area of research. \minife shows a fairly consistent distribution of thread arrival times for the majority of executions; the main opportunity for reclaiming idle time are due to early completion in the one fifth of the time when laggard threads exist. In this case, system periodically transmits all available unsent data with a timeout based on this data would enable threads that were previously idle to efficiently transmit data in these idle times. 

Similarly, the large variance of thread arrival times in \miniqmc and applications similar to it also include significant opportunities for leveraging \earlybird communication. Because this arrival distribution results in 50\% of cores consistently being idle, full applications with workloads similar to \miniqmc (e.g. QMCPACK) would significantly benefit from both a traditional binning model for aggregating data for \earlybird communication and from fine-grain \earlybird communication that does not leverage aggregation.

In contrast, the data shows that \minimd would require a more sophisticated approach to successfully leverage early-bird communication. The first section of \minimd would support a similar timeout or binning-based aggregation approach, but \minimd also includes a second section where this model of overlap is unlikely to succeed. Specifically, most of the threads executed in this section have very similar arrival times, and laggard threads happen in only 4.8\% of our observed iterations. When they do exist, they have high magnitude compared to median run time. Because of this, a more sophisticated approach would need to be used to leverage these relatively rare opportunities for \earlybird communication. 

\section{Related Work}
\label{sec:rel}
In this paper, we examine thread timing measurements in the context of determining the extent to which early-bird communication in partitioned communication may yield faster message delivery.  Thread timing measurements have been used in many existing research efforts, principally as a means of identifying and diagnosing performance issues.  

\paragraph{Profiling Tools}
Several existing tools have been developed to characterize the performance of individual threads in a multithreaded environment.  Mohson et al.~\cite{mohsen2009survey} provides an overview of these tools.  Existing tools for characterizing multithreaded performance use several different approaches for collecting thread execution times, including using the OpenMP profiling interface~\cite{furlinger2005ompp,derose2004profiling,mohr2002design}, specialized processor counters~\cite{intel-vtune}, and information provided by the runtime~\cite{muddukrishna2015characterizing}.  Similarly, Gamblin et al.~\cite{gamblin2008scalable,libra-github} rely on the MPI profiling interface~(PMPI) to characterize per-process execution times in MPI applications.  In this paper, we manually instrument the code in our target workloads to precisely measure per-thread computation time within specific code blocks of interest, \emph{see} \Cref{sec:ins:proxy}.

\paragraph{Performance Analysis}
A common motivation for collecting detailed measurements of thread execution times is to evaluate application performance and to diagnose performance problems (e.g., load imbalance, poor parallelization).  The Performance Optimisation and Productivity~(POP) Centre of Excellence defines a metric based on process execution times for characterizing the extent to which work is evenly distributed across processes.  The Load Balance metric is defined as the ratio of the average process execution time to the maximum process execution time.  Orland and Terboven~\cite{orland2020case} extend this metric to threads and examine load imbalance among OpenMP threads in GMRES.  Muddukrishna et al.~\cite{muddukrishna2016grain} use application characteristics, including thread execution times, to visualize application execution for the purpose of identifying performance bottlenecks.  Liu et al. \cite{liu2013new} break thread execution time into categories that enable them to identify periods of idleness that are indicative of poor performance.  In this paper, we do not explicitly seek to understand the performance of current workloads.  Rather, we use measurements of OpenMP thread execution time to characterize the opportunity to exploit early-bird communication in partitioned communication to deliver message contents earlier.

\paragraph{Fine-Grained Communication}
There have been many other works that have looked fine-grained communication, particularly with an aim to support this form of communication-computation overlap. The gap in knowledge that this paper addresses is how to analyze application behavior to evaluate these techniques. The original partitioned communication paper~\cite{grant2019finepoints} assumes a single laggard thread in the analysis. A notable extension is an evaluation by Temucin et al.~\cite{yiltan2022} who evaluate the performance of partitioned communication under different distributions including a normal distribution. Wombat~\cite{mendygral2017wombat} was focused on adapting a single application to multithreaded communication. Other works that explore fine-grained communication have focused on evaluating the performance of their approach rather than developing a fine-grained application. This includes optimized message matching~\cite{flajslik2016mitigating}
and RMA-MT~\cite{dosanjh2016rma}. Finally, existing work on software offloading~\cite{vaidyanathan2015improving} and MPI Endpoints~\cite{sridharan2014enabling} adapted two computational kernels (QCD Dslash and FFT) to use their fine-grained communication schemes, but they did not evaluate thread timings.

\section{Conclusion}
\label{sec:con}
Fine-grained, thread-safe communication, such as MPI partitioned communication, is a promising approach for enabling efficient communication in multithreaded HPC applications.  In this paper, we instrumented several HPC proxy applications to understand the extent to which \earlybird communication may allow for earlier message delivery.  While additional investigation is necessary, the data we presented in this paper suggests that there may be a meaningful opportunity in scientific HPC applications to achieve better overall communication performance by using \earlybird communication.

\bibliographystyle{plain}
\bibliography{references}
\end{document}